\documentclass[runningheads]{llncs}
\usepackage[T1]{fontenc}

\newcommand{\thiswork}{A\textsc{ff}R\textsc{o}} % name of this work
\newcommand{\db}{A\textsc{ff}R\textsc{o}DB} % name of this work

\newcommand{\precision}{\text{Precision}}

\newcommand{\recall}{\text{Recall}}

\newcommand{\fone}{\text{F}_1\text{-score}}

\newcommand{\accuracy}{\text{Accuracy}}

\newcommand{\thisp}{\text{\thiswork}_\text{P}}

\newcommand{\thisr}{\text{\thiswork}_\text{R}}

\newcommand{\thisf}{\text{\thiswork}_\text{F}}

\usepackage[T1]{fontenc}
\usepackage{xcolor}
\usepackage{subcaption}
\usepackage{algorithm}
\usepackage{algorithmic}
\usepackage{bm}
\usepackage{graphicx}
\usepackage{ulem}
\usepackage{amssymb}
\usepackage{amsmath}
\usepackage{hyperref}
\usepackage{listings}
\usepackage{tcolorbox}
\usepackage{pgfplots}
\usetikzlibrary{pgfplots.groupplots, patterns}
\usepackage{subcaption}
\pgfplotsset{compat=1.18}

\lstdefinelanguage{json}{
    basicstyle=\ttfamily\scriptsize,
    showstringspaces=false,
    breaklines=true,
    frame=none,
    morestring=[b]",
    lineskip=-0.8em,  % Adjust this value for line spacing
}

\begin{document}

% \title{Improving Automatic Mapping of Raw Affiliation Strings to Persistent Organization Identifiers}
\title{From raw affiliations to organization identifiers}

\author{Myrto Kallipoliti\inst{1}\orcidID{0000-0003-2188-6552} 
\and Serafeim Chatzopoulos\inst{2}\orcidID{0000-0003-1714-5225} 
\and Miriam Baglioni\inst{3}\orcidID{0000-0002-2273-9004}
\and Eleni Adamidi\inst{2}\orcidID{0000-0001-9925-1560}
\and Paris Koloveas\inst{2,4}\orcidID{0000-0003-2376-089X}
\and Thanasis Vergoulis\inst{2}\orcidID{0000-0003-0555-4128}
}
\authorrunning{M. Kallipoliti et al.}
\institute{
OpenAIRE AMKE, Athens, Greece\\
\email{myrto.kallipoliti@openaire.eu}\\
\and IMSI, ATHENA RC, Athens, Greece\\
\email{\{schatz, eleni.adamidi, pkoloveas, vergoulis\}@athenarc.gr}\\
\and CNR-ISTI, Pisa, Italy\\
\email{miriam.baglioni@isti.cnr.it}
\and University of the Peloponnese, Tripolis, Greece\\
\email{pkoloveas@uop.gr}
}
\maketitle              % typeset the header of the contribution
\begin{abstract}
Accurate affiliation matching, which links affiliation strings to standardized organization identifiers,
is critical for improving research metadata quality, facilitating comprehensive bibliometric analyses, and supporting data interoperability across scholarly knowledge bases. 
Existing approaches fail to handle the complexity of affiliation strings that often include mentions of multiple organizations or extraneous information.
In this paper, we present 
\thiswork, a novel approach designed to address these challenges, 
leveraging advanced parsing and disambiguation techniques.
We also introduce \db, an expert-curated dataset to systematically evaluate affiliation matching algorithms, ensuring robust benchmarking.
% Our solution is accessible via a public API, allowing seamless integration into third-party workflows. 
Results demonstrate the effectiveness of \thiswork~in accurately identifying organizations from complex affiliation strings.


\end{abstract}

\section{Introduction}

Affiliation information in research publications and other research products (e.g., datasets, project deliverables) is valuable not only for researchers seeking to identify the organizations behind specific research, but also as a rich source of data for generating insights. For example, such information is invaluable for compiling annual reports and statistics for research-performing organizations, or producing university rankings\footnote{E.g., CWTS Leiden Ranking: \url{https://www.leidenranking.com/}}. 

Most research products include information about the authors’ affiliated organizations, typically represented as text strings. These strings often vary in clarity, granularity, structure, and format, making the development of solutions for useful applications a non-trivial task. In recent years, the inclusion of persistent identifiers (PIDs) for organizations, like ROR\footnote{Research Organization Registry~\cite{ror}} IDs, in affiliation metadata has become a popular practice as a way to address this problem. However, even today, a significant number of research products do not include organization PIDs in the affiliation section. 

To alleviate this problem, considerable effort has been devoted by various teams to developing accurate \textit{affiliation matching} approaches, where `raw' affiliation strings are mapped to standardized identifiers~\cite{buttrick2023openalex,kinney2023semantic,s2aff}.  
Such techniques have been used to enrich scholarly knowledge graphs like the OpenAIRE Graph~\cite{openaire_graph} and OpenAlex~\cite{priem2022openalex} making it easier for value-added services that leverage their contents to provide meaningful functionalities for the  research community.

%nd the involvement of diverse organizations in their creation, consistent identification of these institutions has become critical for aggregating and analysing research output, tracking funding trends, and identifying collaborations across different institutions.} 
%{\color{red}Furthermore, affiliation matching enriches metadata used in scholarly knowledge bases (e.g., ), and research discovery portals (e.g., OpenAIRE Explore and BIP! Finder~\cite{Vergoulis2019}), facilitating more efficient literature searches 
%and enabling citation-based impact indicators (such as the the citation index) to be aggregated in the organization level. 
For researchers, funders, and policymakers, affiliation matching enables the tracking of institutional contributions to global challenges and innovation, helping to recognize collaborative research efforts while tracking geographic distribution. 
For instance, when an affiliation string from a publication is matched to a persistent identifier like a ROR ID, it links the data to a comprehensive record of the organization, including several related metadata such as its location, thus adding value to scholarly metadata.
Furthermore, having affiliation data mapped to organization persistent identifiers allows for improved data interoperability across different research databases, which is essential for large-scale data integration and bibliometric analyses. Consequently, since the PID is often missing from the manuscript, mapping affiliation strings to organization PIDs becomes a crucial task.

Affiliation matching is not a straight forward task; naive approaches, like solely exploiting string similarity, are not expected to work perfectly for various reasons. For instance, not all components of affiliation strings hold the same importance when matching them to other affiliation strings (e.g., matching the organization name is more critical than matching the exact address that sometimes is included in the string). 
In addition, affiliation strings often refer to more than one organization: 
the affiliation \textit{`Biomedical Research Institute, Shenzhen Peking University-The Hong Kong University of Science and Technology Medical Center, Shenzhen, China'}, derived from Crossref as the affiliation of a single author,
contains two distinct organizations.
%{\color{red}
%It is evident that the number of organizations mentioned within a single affiliation string is uncertain; therefore, the goal is to accurately identify all such organizations.
%Finally, traditional search methods return a list of results, based only on keyword relevance, without actually determining which, if any, of the returned organizations are actually present in the input affiliation string.
%Consequently, depending solely on a keyword-based search approach for affiliation matching poses significant challenges, considering that
%affiliation matching is often part of a larger data workflow.}
As a result, although various affiliation matching algorithms have been introduced in the literature, there is still space for improvement.
%, especially considering that often an affiliation string includes mentions of multiple organizations.

At the same time, training and evaluating such algorithms depends on access to well-curated ground truth datasets of sufficient size. However, currently, the availability of such datasets remains limited. Existing datasets often suffer from a range of issues, including small scale, a mix of curated and automatically generated entries, inconsistencies in record handling due to varying expert assumptions, lack of transparency regarding the number of annotators of each record and the version of the affiliation database used, and outdated content. These shortcomings undermine the reliability of any experimental results that rely on them as ground truth.

In this work, we introduce \thiswork~(\underline{Aff}iliation strings to \underline{R}esearch \underline{O}rganiza\-tions), a novel affiliation matching approach that aims to address most of the previously described challenges and \db\footnote{\db: \url{https://doi.org/10.5281/zenodo.15322097}}, a publicly available expert-curated dataset to enable the reliable training and systematic evaluation of affiliation matching approaches. In addition, we use \db~along with a third-party dataset of affiliation links to demonstrate the effectiveness of \thiswork~and compare it to state-of-the-art approaches showcasing that our method has superior precision. Finally, we offer an \thiswork~implementation as an open source software\footnote{\thiswork~GIT repository: \url{https://github.com/mkallipo/affiliation-matching}} and we make the data behind \db~available through an open API\footnote{\db~API: \url{https://affro-api.imsi.athenarc.gr/docs}} to offer additional integration options.

It worth noting that a configuration of \thiswork~has already been integrated into the proudction workflow of the OpenAIRE Graph~\cite{openaire_graph}, one of the largest existing scholarly knowledge graphs, enhancing the quality and completeness of its research products metadata. 

%\ser{
%{\color{red}[CAN BE REMOVED]}
%The remainder of this paper is structured as follows: Section \ref{sec:background}~provides background information and defines the problem of affiliation matching. 
%Section \ref{sec:approach}~describes the methodology behind \thiswork, detailing the techniques used for parsing and matching affiliation strings. 
%Section \ref{sec:dataset}~introduces \db, our expert-curated dataset, and outlines its curation process and characteristics. 
%Section \ref{sec:evaluation}~presents the experimental evaluation of our approach, comparing it against existing methods. 
%Section \ref{sec:related}~reviews related work and, 
%finally, Section~\ref{sec:conclusion}~concludes the paper and discusses future research directions.
%}%\input{sections/2-background}
\section{Background \& Problem definition}
\label{sec:background}

In recent years, various initiatives (e.g., ROR~\cite{ror}, GRID~\cite{grid})
%\ser{OpenOrgs~\cite{atzori1openorgs}}
have been engaged in providing persistent identifiers (PIDs) for research-performing organizations. Formally, a PID can be defined as follows.

\begin{definition}[PID]
A persistent identifier (PID) is a long-lasting reference (e.g., a unique string) 
that unambiguously identifies a specific digital entity (e.g., publication, person, organization). 
\end{definition}

PIDs are designed to remain unchanged over time, 
even if the metadata or location of the identified entity changes. 
This ensures stable referencing, and interoperability across different scholarly knowledge bases. 
% {\color{olive} \sout{{\color{blue} A}  main resource that provides persistent identifiers for organizations is the Research Organization Registry (ROR), an open, community-driven initiative that provides unique persistent identifiers for a wide range of research organizations worldwide. 
% ROR identifiers (i.e., ROR IDs) are the de facto standard for uniquely identifying research organizations in scholarly data,
% and point to structured entries for organizations.}
% Both these resources, enable affiliation strings to be linked to specific, standardized organizational entries, assisting data interlinking and fostering improved data quality.
% }

The process of mapping raw affiliation strings, represented in text, to organization persistent identifiers is called \textit{affiliation matching} and can be formally defined as follows:

\begin{definition}[Affiliation matching]
Let $P$ be a set of distinct organizational persistent identifiers. Given a raw affiliation string $s$, affiliation matching is the process of finding the maximal subset $P_s$ of identifiers $\varnothing \subseteq P_s \subseteq  P$, such that the elements of $P_s$ correspond to the organizations referenced in $s$.
\end{definition}

It is worth mentioning that the definition above accommodates cases where affiliation strings contain multiple organizations, a scenario that is frequently observed in practice (but overlooked by various approaches).
A key step to effectively identify and link these organizations is measuring textual similarity:
in this context, similarity measures are a core component 
of affiliation matching algorithms, as they quantify the degree of resemblance between raw affiliation strings and organization names.
A well-known such similarity measure, that also allows for effective ranking of potential matches is \textit{cosine similarity}, that can be defined as follows:
% In our approach, we employ \textit{cosine similarity} as a core component to measure the similarity between the raw affiliation strings and the names of organizations; 

\begin{definition}[Cosine similarity]
\label{def:cosine}
Given two vectors $\vec{A}$ and $\vec{B}$, cosine similarity is defined as:
\[
\text{sim}(\vec{A}, \vec{B}) = \frac{\vec{A} \cdot \vec{B}}{\|\vec{A}\| \|\vec{B}\|}
\]
where $\vec{A} \cdot \vec{B}$ is the dot product of $\vec{A}$ and $\vec{B}$, and $\|\vec{A}\|$ and $\|\vec{B}\|$ are the magnitudes of $\vec{A}$ and $\vec{B}$, respectively. The result is a value in the range $[0, 1]$, where $1$ indicates perfect similarity and $0$ indicates no similarity.
\end{definition}

In the context of affiliation matching, vectors $\vec{A}$ and $\vec{B}$, represent textual features derived from affiliation strings and organization names (e.g., word frequency representations or token embeddings). 
Cosine similarity allows us to measure the semantic and/or syntactic closeness between raw affiliation strings and organization names,
as well as rank potential matches effectively.
% \sout{
% . By leveraging this measure, potential matches can be identified and ranked effectively. Additionally, when combined with advanced parsing and disambiguation techniques, it significantly improves the accuracy of detecting organizations within complex affiliation strings.
% }

\section{Methodology}
\label{sec:approach}

In this section, we present \thiswork, our affiliation matching approach. Section~\ref{sec:app-overview} offers a high-level overview of the approach, while Sections~\ref{sec:preprocessing}-\ref{sec:finalization} delve into the technical details of its main phases. 

\subsection{High-level overview}
\label{sec:app-overview}

\begin{figure}[h]
    \centering
    \includegraphics[width=1\linewidth]{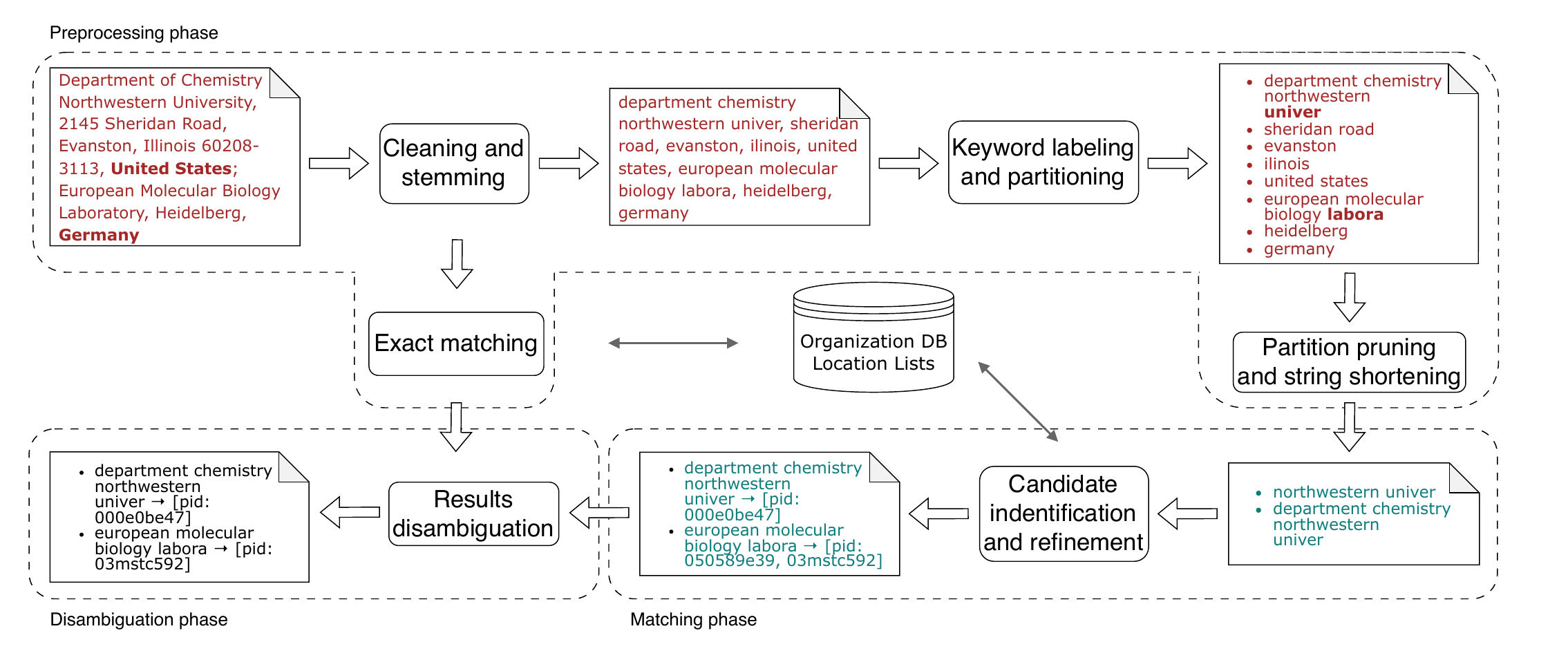}
    \vspace{-0.7cm}
    \caption{Example of \thiswork's key processing steps.}
    \label{fig:affro-steps}
\end{figure}

Figure~\ref{fig:affro-steps} illustrates the main stages of \thiswork~and offers a running example based on an indicative input affiliation string. \thiswork~is built on a rule-based, keyword-driven framework that can be divided into three main phases: 
\begin{enumerate}
    \item \textit{Preprocessing}, where the input affiliation strings are tokenized, formatted, and cleaned.  
    \item \textit{Matching}, where the extracted tokens are matched with records into a data\-base of research-related organizations. 
    \item \textit{Disambiguation}, where vague matches are further refined, if needed. 
\end{enumerate}
 
It is worth noting that our approach is applicable to any database of organizations that provides persistent identifiers (PIDs) and country information. Additionally, it can leverage the city of organizations and any parent-child relationships among them as optional fields; while not required, this information can improve performance. 
% when available. 
In our work, we have used the ROR database~\cite{ror}, 
therefore ROR IDs are as used as identifiers in our curated dataset (see Section~\ref{sec:dataset}).  

Of course, before the database can be utilized by the approach, a preprocessing phase is necessary to clean and format the affiliation strings in a manner that facilitates matching, based on the cleaning and stemming techniques described in Section~\ref{sec:clean}, and the pruning method outlined in ~\ref{sec:filter}. This phase also involves constructing a series of dictionaries to organize the information in a convenient way for the various phases of \thiswork. The main dictionaries are described in the respective parts of the approach that are used (in Sections~\ref{sec:preprocessing}-\ref{sec:finalization}).

Finally, it is worth mentioning that our approach has four parameters (namely \texttt{window}, \texttt{sim\_u}, \texttt{sim\_o}, and \texttt{specific}) that affect its accuracy. These parameters and their effects are elaborated in the following sections.

\subsection{Preprocessing phase}
\label{sec:preprocessing}

The preprocessing phase of \thiswork~consists of a series of steps outlined in the following sections. %Sections~\ref{sec:clean}--\ref{sec:short}. 
%Section~\ref{sec:org_dict} includes the description of dictionaries to catalog PIDs and organizational metadata.

\subsubsection{Cleaning and stemming.}
\label{sec:clean}

% As a first step, raw affiliation strings undergo a cleaning process to ensure a standardized format before mapping. This is a technical task developed after reviewing thousands of raw affiliation strings. It involves multiple stages including:

% \begin{itemize}
%     \item  Inserting a space between consecutive letters where a lowercase letter is followed by an uppercase one (e.g., aB → a B) and  subsequently converting the text to lowercase.
%     \item Ensuring consistent character encoding.
%     \item Removing stopwords, special characters, and multi-digit numbers, as well as potential leading numbers in the string.
%     \item Removing parentheses that do not contain key terms such as \texttt{"University"}, \texttt{"Hospital"}, or \texttt{"Clinic"}.
%     \item Replacing double consonants with a single one.
%     \item Standardizing abbreviations such as \texttt{"St."} for \texttt{"Saint"} and \texttt{"Inst."} for \texttt{"Institute"}
%     \item Addressing potential typographical errors, specifically concerning the term \texttt{"University"}. 
%     \item Converting Roman numerals (e.g. \texttt{i}, \texttt{ii}, \texttt{iii}) to their numeric equivalents (e.g., \texttt{1}, \texttt{2}, \texttt{3}).
% \end{itemize}

This step implements a series of targeted preprocessing actions tailored to improve the effectiveness of the matching process (see Section~\ref{sec:matching}). These include lowercasing the strings and removing stopwords, special characters, and multi-digit numbers—including potential leading numerals. 
Parentheses that do not enclose key institutional terms such as \texttt{``University''}, \texttt{``Hospital''}, or \texttt{``Clinic''} are also removed. Additional normalization involves collapsing double consonants into a single instance, standardizing common abbreviations (e.g., \texttt{``St.''} to \texttt{``Saint''}, \texttt{``Inst.''} to \texttt{``Institute''}), performing some language processing to facilitate stemming\footnote{For instance, in some special cases, translation of critical terms is applied. As an example, German terms such as \texttt{``Universitätsklinikum''} and \texttt{``Universitätsbibliothek''} are replaced by their corresponding English equivalents.} and correcting typographical errors related to critical terms such as \texttt{``University''}. Finally, we convert Roman numerals (e.g., i, ii, iii) to their corresponding numeric forms (1, 2, 3) to enhance downstream matching accuracy.

Next, stemming is applied to facilitate string comparisons by reducing words to a common root (e.g., \texttt{``University''}, \texttt{``Universität''}, and \texttt{``Université''} are all stemmed into \texttt{``univer''}). 
Finally, any country names detected in the raw affiliation string are stored in a dictionary for later use in narrowing the candidate set during the matching phase.
% As shown in Figure~\ref{fig:affro-steps}, the output of the cleaning and stemming step for the input affiliation string in the running example is: \texttt{"department chemistry, northwestern univer, sheridan road, evanston, ilinois, united states; department nanoengineering univer california san diego la jola ca usa"}

\subsubsection{Keyword labeling, partitioning, and exact matching.}
\label{sec:tokens}

In this step, the algorithm first checks whether the cleaned and stemmed affiliation string exists in the database of organizations. If so, it is passed directly to the disambiguation phase (Section~\ref{sec:finalization}) implementing a shortcut. If this is not the case, the keywords of each affiliation string are classified into categories (that are used by parts of the approach) and, then, the affiliation strings are partitioned to facilitate the matching process.
%Then, algorithm checks whether exact matches for the partitions exist in the database of organizations. If at least a match is found, the partition is passed directly to the disambiguation phase (Section~\ref{sec:finalization}) implementing a shortcut. 

Keyword classification assigns each keyword to one of the following categories: 
\begin{itemize}
    \item \texttt{``basic\_key''}: keywords representing basic types of research organizations (i.e., \texttt{``univer''}, \texttt{``institu''}, \texttt{``hospital''}, and \texttt{``labora''}).
%    \item \texttt{``add\_cconj''}: the additive coordinating conjuction, i.e., the "and" term
    \item  \texttt{``country''}: keywords included in a dictionary of countries that we have created based on the database of organizations. 
    \item \texttt{``other''}: keywords that do not belong to the other categories.
\end{itemize} 

\thiswork~partitions the input affiliation string based on a predetermined set of delimiting characters (at the point of writing, a comma ``$,$'', a semicolon ``$;$'', a colon ``$:$'', a forward slash ``$/$'', and an en dash ``$-$'').
%\footnote{{\color{red}Based on the organization data we may need to exclude a few strings from this process. For example, in the ROR case, some common names of universities located in different cities, like "University of California, Los Angeles" or "Islamic Azad University, Babol" are not tokenized.}} 
The generated partitions are further divided as follows: If the word \texttt{``and''} appears between two basic keywords, it is removed, and the partition is split at that position.
%Note that the \texttt{country} label will be utilized later for verification purposes. 
Additionally, a partition that cannot be matched in the organization database and contains the preposition \texttt{"at"}, is split at its position (and the new partitions are checked again for matches in the database).

% Returning to the running example, the set of tokens generated for the input string is:  \{$\texttt{``department chemistry", ``northwestern univer", ``sheridan road", ``evanston",} \\ \texttt{``ilinois", ``united states", ``department nanoengineering univer california} \\ \texttt{san diego la jola ca usa"} \}$. 

\subsubsection{Partition pruning.}
\label{sec:filter}

After partitioning, \thiswork~applies certain rules to reduce the set of partitions. The  objective is to concentrate comparisons on the parts of the affiliation string that are more crucial for effective matching while reducing computational demands. 
%The rules were determined based on the results of a statistical analysis conducted on the ROR and OpenAIRE database records, with the possibility of further adjustments in future versions of the approach. 
Currently, the following rules are applied:
\begin{itemize}  
    \item \textit{Keyword filter.} A dictionary of frequent keywords used in names of research institutions (e.g.,\texttt{``univer''}, \texttt{``institu''}, \texttt{``academ''}, \texttt{``school''}, \texttt{``lab''}, \texttt{``clinic''},  \texttt{``library''}) has been created based on a statistical analysis of the ROR database. 
    When the $\texttt{specific}$ parameter is set to $\texttt{True}$, the list is extended with frequent acronyms (e.g., \texttt{``riken''}, \texttt{``eth zurich''}, \texttt{``cnrs''}, \texttt{``ucla''}) and entity names (usualy companies e.g., \texttt{``roche''}, \texttt{``astrazeneca''}, \texttt{``ibm''}, \texttt{``toyota''}) retrieved from the Nature Index.\footnote{Nature Index: \url{https://www.nature.com/nature-index/}} \thiswork~filters out partitions that do not contain any of these keywords. 
    \item \textit{Banned partitions.} \thiswork~does not consider partitions that are generic terms such as \texttt{``univer''} or \texttt{``clinic''} if no city is present in a following partition of the same string,\footnote{{\scriptsize \url{https://github.com/mkallipo/affiliation-matching/blob/main/txt_files/city_names.txt}}}
    \footnote{{\scriptsize\url{https://github.com/mkallipo/affiliation-matching/blob/main/txt_files/banned.txt}}} as well as address-related phrases like \texttt{``univer road''} or \texttt{``univer str''}. In addition, partitions that do not contain a basic keyword and contain informative  substrings about authors' placements, such as \texttt{``pro\-fessor''}, \texttt{``assistant''}, and \texttt{``phd''}, as well as substrings referring to child organizations like \texttt{``program''}, \texttt{``unit''}, and \texttt{``department''} are also disregarded. The latter is due to the fact that the ROR database statistics indicate that such child organizations typically do not receive a dedicated PID.\footnote{For instance, we have found that only approximately $0.06\%$ of ROR records correspond to departments that are children of universities.} This rule facilitates mapping the string to the parent organization’s PID.
    The precise list of such rules can be found in our code repository.\footnote{{\scriptsize\url{https://github.com/mkallipo/affiliation-matching/blob/main/create_input.py}}}
\end{itemize}

\subsubsection{String shortening for universities and entity names.}
\label{sec:short}

In this final step before matching, \thiswork~narrows down the remaining partitions which are not in the organization database, by preserving words near the substring \texttt{``univer''} using the \texttt{window} parameter. 
This parameter gets positive integer values and is used to trim long university-related strings. Any terms that are outside the \texttt{window} are discarded. 
Additionally, if a token contains a keyword from the \texttt{Entities} category (see Section~\ref{sec:filter}), the token is reduced to the keyword. 

% The set of tokens of our running example becomes: $\{$\texttt{"northwestern univer", "department nanoengineering univer california san diego"}$\}$.

%As a result, the average string length is reduced from $\sim$90 to $\sim$35 characters.

% \noindent\textbf{Examples}
% \begin{itemize} 
%     \item $s_1=$ \texttt{"Cariology and Operative Dentistry, Graduate School of Medical and Dental Sciences, Tokyo Medical and Dental University"}, with  $\mbox{length}(s_1) = 118$.
%     After tokenization and removal of stop-words, $s_1$ becomes \texttt{"cariology operative dentistry, graduate school medical dental sciences, tokyo medical dental university"} and finally $\bar{s}_1=$  \texttt{"tokyo medical dental university"}, with $\mbox{length}(\bar{s}_1)=31$.
    
%     \item Let $s_2=$ \texttt{"Centre for Epidemiology and Biostatistics, Melbourne School of Population and Global HealthThe University of Melbourne Melbourne VIC Australia"}, with $\mbox{length}(s_2) = 142$. 
%     After tokenization and removal of stop-words, $s_2$ becomes 
%     \texttt{"centre epidemiology biostatistics, melbourne school population global healththe university melbourne melbourne vic australia"}, and then \texttt{"melbourne school population global healththe university melbourne melbourne vic australia"}. Finally, we shorten further the string by keeping only the words lying at most three words away from \texttt{"university"}, finally obtaining  $\bar{s}_2=$ \texttt{"population global healththe university melbourne melbourne vic"}, with $\mbox{length}(\bar{s}_2)= 62$.
% \end{itemize}

\subsection{Matching phase}
\label{sec:matching}
After the completion of the preprocessing, \thiswork~is ready to search for matches of the final partitions inside the organization database. The respective process consists of two steps: first, the identification of candidate matches (Section~\ref{sec:cand}) and, then, the refinement of the results (Section~\ref{sec:refinement}). 

\subsubsection{Candidate identification.}
\label{sec:cand}
At this stage, the algorithm employs cosine similarity to determine whether input partitions (approximately) match names in the organization database. 
As described in Section~\ref{sec:in-depth-param-analysis}, experimentation revealed that optimal similarity thresholds vary across organization types. Specifically, for universities, a looser threshold, denoted as $\texttt{sim\_u}$, proves to be more effective compared to other types of organizations, for which a stricter threshold, $\texttt{sim\_o}$, is applied.
%Additionally, the parameter \texttt{limit} is used to halt the matching process if the number of possible matches for a given token exceeds the set threshold.}
In Section~\ref{sec:eval-config}, we provide insights on the optimal selection of values for those parameters. 
Note that the algorithm searches for possible matches not across all organization names, but only within the subset corresponding to the countries mentioned in the raw affiliation string. This is a crucial step that improves both the accuracy and performance of the algorithm.
%As it will be evident in Section~\ref{sec:eval-config}, a good choice for these thresholds is  \texttt{SimU} $\sim$ 0.6 and  \texttt{SimG} $\sim$ 0.85. Alternative methods like Levenshtein Distance or Jaro-Winkler Distance were considered for measuring string similarity, and it was concluded that cosine similarity is the most appropriate choice for this specific application.

\subsubsection{Results refinement.}
\label{sec:refinement}

If the algorithm detects multiple matches (candidates) above the specified similarity thresholds for a token, it proceeds with an additional check to determine the best fit among these candidates. 
To this end, cosine similarity is applied once again, comparing the candidates with the original cleaned and stemmed affiliation string.
This comparison is meaningful because it takes into account additional information from the original affiliation, such as city names or the names of child organizations. The algorithm aims to identify the best match, additionally incorporating Levenshtein distance at this stage to account for potential typos. For the final refinement, the number of occurrences of basic keywords is also considered to limit the number of results.

\subsection{Results disambiguation}
\label{sec:finalization}

Once the final match is identified, the algorithm checks whether the organization's name corresponds to multiple entries in the database. 
If so, the algorithm leverages available city and country information to determine the most suitable match. 
If no location is provided and the organization's name falls into a special category of known entities (e.g., tech companies), then the algorithm matches to the parent organization if existent (for example, in case of ROR IDs, the affiliation "Google" is mapped to \texttt{"00njsd438"}). 
If none of these conditions are met, then a match is returned only if there is a unique active ID among the possible results the name corresponds to a unique active ID.  
% See also Figure~\ref{fig:final_candidate}.
Finally, if the identified organization has a non-active ID, the IDs of any successor organizations are retrieved and returned.

%\begin{figure}[h]
%     \centering
%     \includegraphics[width=1\textwidth]{plots/final_candidate.pdf}
%     \begin{center}
%     \caption{Path from final candidate to result {\color{red}[REMOVE?]}}
%     \label{fig:final_candidate}
%     \end{center}
% \end{figure}

% Returing to our running example, the tokens to match are:  
% \begin{itemize}
%     \item $t_1$ = \texttt{"northwestern univer"}, which is matched to \texttt{https://ror.org/000e0be47} with similarity score $1$.
%     \item $t_2$ = \texttt{"department nanoengineering univer california san diego"}, which is matched to \texttt{https://ror.org/0168r3w48} with similarity score $0.82$.
% \end{itemize}

\section{Expert-curated dataset}
\label{sec:dataset}

In this section, we describe \thiswork DB, the curated dataset that we have compiled to evaluate the accuracy of affiliation matching approaches. 
%Unlike similar datasets (e.g., OpenAlex Test Dataset~\cite{openalex_test_dataset}), which rely on (semi-)automated approaches for creation, our dataset is meticulously curated entirely by experts, ensuring superior accuracy and reliability. 
In the following, we outline the process of creating \db~ (Section~\ref{sec:data-creation-process}), the structure and format of the data (Section~\ref{sec:data-structure}), as well as its update policy (Section~\ref{sec:data-update-policy}).

\subsection{Dataset creation process}
\label{sec:data-creation-process}

In the following, we provide an overview of the methodology employed in compiling the dataset, highlighting the annotation process to ensure data quality and consistency.
Each entry in the dataset has been independently annotated by at least two experts to improve consistency and reduce bias. Furthermore, the experts were instructed to identify multiple organizations when applicable and to include detailed metadata capturing the nature of relationships between these organizations. Because of the previous reasons, our dataset serves as a robust and comprehensive benchmark for evaluating affiliation matching algorithms.

The expert-driven process for creating the dataset was carefully designed. 
We first randomly selected approximately $1,500$ records from Crossref, with non-empty affiliation lists. 
Each affiliation string was evaluated by two experts (we have used five experts in total)
who identified the corresponding ROR entries it represented, and then classified each link into one of three categories: \texttt{EXACT}, \texttt{ANCESTOR}, and \texttt{VAGUE}.
Table \ref{tbl:annotation-labels} summarizes these annotation categories along with their interpretation.
In cases of disagreement between the two experts, a third expert reviewed the affiliation string and made the final decision on
the labels and corresponding links.

\begin{table}[h]
\caption{Annotation categories for expert-curated affiliation links.}
\label{tbl:annotation-labels}
\centering
\begin{tabular}{|l|l|}
\hline
\textbf{Name} & \textbf{Description} \\ 
\hline
\texttt{EXACT}     & 
\parbox[t]{10.5cm}{The ROR ID found corresponded precisely to the most specific organization mentioned in the string.}\\ 
\texttt{ANCESTOR}  & 
\parbox[t]{10.5cm}{The ROR ID corresponded to a parent organization mentioned in the string.} \\ 
\texttt{VAGUE}     & 
\parbox[t]{10.5cm}{No elevant ROR ID was detected or the expert could not confidently assign a corresponding ROR ID.}\\
\hline
\end{tabular}
\end{table}

% who had to:
%
% \begin{itemize}
%     \item Label the string with one of the following \footnote{All links mentioned are \textbf{active} ROR IDs}:
%     \item Provide the corresponding ROR links and assign to each of them the label "exact" or "ancestor".
% \end{itemize}

The final dataset was constructed after
excluding
affiliation links labeled as \texttt{VAGUE};
% Table~\ref{tbl:dataset-stats}~summarizes some statistics regarding this dataset.
In particular, it consists of $1,374$ raw affiliation strings, extracted from $1,261$ DOIs.
Notably, the total number of ROR links is $1,475$, 
since experts assigned multiple ROR links to some records,
$379$ of them corresponding to 
exact matches, and $1,096$ to matches to
ancestor organizations.
A third expert was enlisted to resolve disagreements in the evaluation of 110 raw affiliation strings, ensuring consensus and accuracy in the final assessments.
% mentioned in the string), 
% as some records have multiple ROR links assigned by the experts. 

% Finally, for 110 raw affiliation strings, a third expert was required to finalize the evaluation. 

% \begin{table}[h]
% \caption{Summary statistics for the expert-curated dataset.
% {\color{red}CAN BE REMOVED}}
% \label{tbl:dataset-stats}
% \centering
% \begin{tabular}{|l|c|}
% \hline
% \textbf{Statistic} & \textbf{\#} \\ 
% \hline
% Number of DOIs & 1,261 \\ 
% Number of raw affiliation strings & 1,374 \\ 
% Total ROR IDs & 1,484 \\ 
% Distinct ROR IDs & 1,048 \\
% Exact matches & 381 \\ 
% Ancestor matches & 1,103 \\ 
% \hline
% \end{tabular}
% \end{table}

\subsection{Data structure description}
\label{sec:data-structure}

The dataset is formatted as a JSON Lines (JSONL)\footnote{JSON Lines format documentation: \url{https://jsonlines.org/}.}
file where each line contains a valid JSON object.
This format is rather simple yet efficient for large datasets, since it facilitates incremental processing of the data.  
% An indicative record is depicted in Figure~\ref{fig:data-structure}. 
The expert annotations are captured by instances of an \texttt{annotation object} which contains two fields: (a) \texttt{exact}, and (b) \texttt{ancestor}, which correspond to arrays of ROR identifiers identified as exact or ancestor matches respectively. Then,
each JSON object contains the following four fields:
\begin{enumerate}
    \item \texttt{raw\_affiliation\_string}: a string representing the raw affiliation data.
    \item \texttt{extracted\_dois}: the DOIs from which the affiliation string was extracted.
    \item \texttt{expert\_judgements}: expert annotations each with the following fields:
        \begin{enumerate}
            \item \texttt{expert\_id}: a positive integer identifying the expert.
            \item \texttt{matches}: an \texttt{annotation object} representing the expert's annotation.
        \end{enumerate}
    \item \texttt{final\_judgement}: an \texttt{annotation object} representing the final evaluation.

\end{enumerate}

% The number of raw affiliation strings reviewed by each expert is as follows:
% \begin{itemize}
%     \item expert\_1: 816
%     \item expert\_2: 755
%     \item expert\_3: 196
%     \item expert\_4: 162
%     \item expert\_5: 929
% \end{itemize}

% \subsection{Examples}
% Two examples of data records are provided below.
% {\color{red}[TODO: include examples in figures - important to reduce space]}

% \begin{figure}[h]
%     \centering
%     \begin{tcolorbox}[colframe=black!50, colback=white, arc=3pt, boxrule=0.5mm, width=\textwidth]
%         \begin{lstlisting}[language=json]
% {
%     "raw_affiliation_string": "1 Department of Biochemistry, Molecular Biology and Cell Biology, Northwestern University, Evanston, Il 60208-3500, USA",
%     "extracted_dois": [
%         "10.1242/dev.120.9.2673"
%     ],
%     "experts_judgements": [ {
%             "expert_id": "1",
%             "matches": {
%                 "exact": [],
%                 "ancestor": [
%                     "https://ror.org/000e0be47"
%                 ]
%             }
%         }, {
%             "expert_id": "2",
%             "matches": {
%                 "exact": [],
%                 "ancestor": [
%                     "https://ror.org/00m6w7z96"
%                 ]
%             }
%         },
%         ...
%     ],
%     "final_judgment": {
%         "exact": [],
%         "ancestor": [
%             "https://ror.org/000e0be47"
%         ]
%     }
% }
%         \end{lstlisting}
%     \end{tcolorbox}
%     \caption{Structure of an indicative record from the expert-curated dataset.}
%     \label{fig:data-structure}
% \end{figure}

Our dataset is openly available on Zenodo\footnote{\thiswork DB dataset: \url{https://doi.org/10.5281/zenodo.15322097}} under the CC0 Creative Commons license, ensuring unrestricted access and enabling broader reuse. Moreover, we plan to provide regular updates by incorporating new entries and revising existing ones based on the latest information available in the ROR database. 

\subsection{Update policy}
\label{sec:data-update-policy}

Our goal is for \db~to serve as a living dataset that will be regularly updated  through a carefully designed process that ensures the inclusion of matches for new affiliation strings, as well as the revision of existing entries to correct errors or to add matches to ROR IDs that were not available at the time the initial records were created. This update process is intended to keep the dataset accurate and useful over time, while also facilitate the involvement of additional experts in the curation process.

With each new release of the dataset, a batch of affiliation strings will be provided for expert curation. This batch will be selected to ensure it includes strings from the following categories:  
\begin{enumerate}
    \item \textbf{New affiliation strings:} a random set of new affiliation strings from Crossref consisting of approximately $70\%$ of the records.
    \item \textbf{Old affiliation strings:} a random subset of the affiliation strings that have been processed by the experts of a previous dataset version. Both strings with non-empty and empty exact or ancestor matches will be selected so that we can both validate the accuracy of assigned identifiers and identify missing ones. 
\end{enumerate}

Additionally, a fully automated process will take place searching inside the old dataset records with the aim to identify ROR IDs that are no longer active. In case a successor organization is determined by ROR, the inactive identifier will be replaced by its identifier. Otherwise, the identifier, will be removed.
  
   % \item \textbf{Validating existing affiliations:} A random subset of … of affiliations with non-empty exact or ancestor lists is selected for expert review to verify the accuracy of the assigned ROR IDs and identify any missing ones.

    %\item \textbf{Reviewing unmatched affiliations:} A random subset of … of affiliations with empty exact or ancestor lists is selected for review to determine if new ROR IDs can be assigned.

    %\item \textbf{Incorporating new affiliations:} A random set of … new raw affiliation strings from Crossref is processed using the same method as in the preparation step to enrich the dataset.

\subsection{Open API}

Researchers and developers can leverage the affiliation resolution functionality of \thiswork\ through a publicly accessible API to enrich publication metadata in their research workflows.
The main endpoint of this API takes as input a raw affiliation string, in a form typically found in publication metadata, along with \thiswork's configuration parameters (see Section~\ref{sec:app-overview}), 
and returns a ranked list of matched research organizations. 
Each result includes a unique identifier (typically a ROR ID), 
and a numerical confidence score. 
This API offers a convenient way to experiment with the results of \thiswork\ out of the box, while for more demanding use cases, 
the service is also available as open source,\footnote{\thiswork\ API GIT repository: \url{https://github.com/mkallipo/affiliation-matching-api}} allowing deployment on self-managed environments.

\subsection{Comparison with other datasets}
\label{sec:other-datasets}

Unlike other similar datasets, ours includes affiliation strings that mention multiple organizations, providing corresponding links to multiple identifiers.  
Furthermore, most well-established datasets with affiliation links (see Section~\ref{sec:related}) combine curated and automatically generated links, which can undermine the reliability of experiments using them due to differing levels of trust between the two types of records. In contrast, \db~is a fully expert-curated dataset.

In addition, \db~prioritizes transparency, consistency, and long-term reliability. Many other datasets do not provide crucial information, such as which version of the ROR dataset was used during the expert annotation process. This is an important detail, given that ROR is continuously evolving and this absence can result in discrepancies where identifiers used in the training of an approach may not have existed at the time of the creation of the expert-curated dataset. Additionally, there is often no indication of how many experts assessed each record, whether multiple experts were involved, or what individual scores were assigned by each expert. To the best of our knowledge, \db~is currently the only expert-curated dataset offering these features.

Moreover, our dataset is built upon a clear and well-defined annotation policy that was shared with all experts to minimize inconsistencies—for example, in handling matches to ancestor organizations. Also, every entry has been reviewed by multiple experts, resulting in high-quality ground truth records. In preliminary evaluations of curated entries from some well-established datasets~\cite{s2aff,kinney2023semantic,openalex_test_dataset}, we identified a noticeable number of errors or missing links. These issues led us to exclude those datasets from our experiments due to trustworthiness concerns.

Finally, a key strength of our dataset is that we have defined a robust update policy for it to ensure the dataset remains current and consistent over time.

% \begin{lstlisting}
% {
%     "raw_affiliation_string": "3Key Laboratory of Breast Cancer in Shanghai, Fudan University Shanghai Cancer Center, Fudan University, Shanghai, China.",
%     "extracted_dois": [
%       "10.1158/2326-6066.cir-21-1072"
%     ],
%     "experts_judgements": [
%       {
%         "expert_id": "1",
%         "matches": {
%           "exact": [
%             "https://ror.org/00my25942"
%           ],
%           "ancestor": []
%         }
%       },
%       {
%         "expert_id": "2",
%         "matches": {
%           "exact": [],
%           "ancestor": [
%             "https://ror.org/00my25942",
%             "https://ror.org/013q1eq08"
%           ]
%         }
%       },
%       {
%         "expert_id": "5",
%         "mathces": {
%           "exact": [],
%           "ancestor": [
%             "https://ror.org/00my25942",
%             "https://ror.org/013q1eq08"
%           ]
%         }
%       }
%     ],
%     "final_judgment": {
%       "exact": [],
%       "ancestor": [
%         "https://ror.org/00my25942",
%         "https://ror.org/013q1eq08"
%       ]
%     }
%   }
% \end{lstlisting}

\section{Experimental evaluation}
\label{sec:evaluation}

In this section, we present the experiments we conducted to evaluate the effectiveness of our approach. 
In Section \ref{sec:setup}, we outline the experimental setup, while in Section~\ref{sec:eval-config} we present our experiments to configure and study the parameters of our approach.
Finally, in Section~\ref{sec:eval-competitors} we outline our findings regarding the effectiveness of \thiswork~in comparison to the state-of-the-art approaches.

\subsection{Setup}
\label{sec:setup}

\subsubsection{Methods.}
\label{sec:methods}

The evaluation involved experiments with three main approaches;  
different configurations were tested for each one, 
where applicable.

\begin{itemize}
    \item \textit{\thiswork}, our proposed approach as presented in Section~\ref{sec:approach}, with three different parameter configurations, each maximizing precision, recall or $F_1$-score (based on the results of the experiments in Section~\ref{sec:eval-config}), respectively. 
    
    \item \textit{S2AFF}~\cite{s2aff,kinney2023semantic}, the approach developed by Semantic Scholar to link affiliation strings with affiliations in ROR. 
    % We consider two variants: $S2AFF_{first}$, which uses only the top-ranked result, and $S2AFF_{all}$, which includes all retrieved results.}
    % \item \textit{ROR}, the approach behind the affiliation matching service provided by ROR's main API (it can be used by setting the `affiliation' query parameter in the `/organizations' endpoint\footnote{Affiliation matching request using ROR API: \url{https://api.ror.org/organizations?affiliation=[raw-affiliation-string]}}), specifically designed to facilitate the  identification of research organizations by mapping free-text affiliation strings to ROR identifiers. While using the API, we only consider high-confidence matches, i.e., those marked with the property `chosen=true' in the response, as suggested by the ROR documentation.\footnote{Documentation for affiliation matching in the ROR API: \url{https://ror.readme.io/docs/matching\#affiliation-parameter-approach}} {\color{blue} IDs that where from later version of the ROR dump where omitted}
    \item \textit{OpenAlex  ROR Predictor}~\cite{buttrick2023openalex}, a modified version of OpenAlex's institution parsing service~\cite{openalex_institution_parsing,priem2022openalex}, extended to also return ROR IDs from affiliation strings alongside OpenAlex institution IDs. 
\end{itemize}

\subsubsection{Datasets.}

For our experiments, we used the following sets of data:
\begin{itemize}
    \item \textit{\db.}
    Our expert-curated dataset, consisting of raw affiliation strings manually linked to ROR IDs (see Section~\ref{sec:dataset} for details).
    % Our expert-curated dataset that contains raw affiliation strings linked to ROR IDs of January 2024 dump, through an expert curation process. 
    % {\color{red}These links are further classified into `exact' and `ancestor' matches.} 
    % This dataset serves as the primary benchmark for the evaluation in this section.

    % \item \textit{ROR database dump.}
    % The Research Organization Registry (ROR) serves as an input to our algorithm,  offering a comprehensive collection of organizational names and identifiers to facilitate accurate matching.
    % \ser{In particular, for our experiments we used the ROR database dump of January 2024.}

    \item \textit{Crossref organization assertions.}
    This dataset contains organization relationships from Crossref metadata, including both publisher-submitted and automatically assigned ROR IDs. 
    We filtered the dataset to retain only records where the ROR ID was submitted by publishers (to avoid automatically matched strings), resulting in $47,942$ unique affiliation strings, each mapped to a single ROR ID.
    % \sout{, even though multiple organizations were present, example "Institute for Experimental Medical Research, University of Oslo and Oslo University Hospital, Oslo, Norway", or "University of Aberdeen \& Durham University"}. 
    % {\color{red}
    % The ROR records, were offline cleaned and stemmed (see aslo Section~\ref{sec:clean}), while the list of important terms (see Section~\ref{sec:filter}) was used to classify all them into organization types removing records of particular type}. 

\end{itemize}

We should note that all the compared approaches were configured to use the ROR database dump of January 2024 (v1.40), since this was the version used by the expert curators of \db~to produce their annotations.

\subsubsection{Evaluation metrics.}
\label{sec:evaluation-measures}
% {\color{blue}There are several ways to evaluate an algorithm's performance, including micro and macro precision, recall, F1-score, and exact match accuracy. For affiliation matching, where one affiliation can have multiple ROR IDs, micro metrics are the best choice because, unlike other methods, they treat each ROR ID equally, making the evaluation fair even when some affiliations have many ROR IDs while others have few. Micro metrics are calculated by first aggregating the contributions of all classes (in this case, ROR IDs) before computing the final result. The formulas for micro precision, recall, and F1-score are as follows:}
To evaluate the effectiveness of the considered approaches, we use micro-averaged \precision, \recall, and $\fone$, which are particularly well-suited for multi-class classification tasks such as affiliation matching. 
In this task, a single affiliation string may correspond to multiple organization identifiers, and the number of these identifiers can vary significantly across affiliations. Micro-averaging addresses this imbalance by aggregating true positives, false positives, and false negatives across all instances before computing the metrics, thereby treating each prediction equally regardless of class frequency.

% \begin{align*}
% \text{Precision} = \frac{TP}{TP + FP}, & \quad \quad 
% \text{Recall} = \frac{TP}{TP + FN}, \\
% F_1 =\frac{2 \cdot \text{Precision} \cdot \text{Recall}}{\text{Precision} + \text{Recall}}
% \end{align*}

% \noindent where $TP$ is the number of true positives (correctly predicted positive samples), $FP$ is the number of false positives (incorrectly predicted positive samples), and $FN$ is the number of false negatives (positive samples missed by the model). 

% However, this common form, is used to evaluate binary classification problems.
% Since affiliation matching is a multi-class classification task, we use the modified versions of these measures, tailored for this scenario.
% In this context, micro-average precision, recall, and $F_1$-score (denoted as $\precision$, $\recall$, and $\fone$, respectively) are more appropriate since they aggregate the contributions of true positives, false positives, and false negatives across all classes (affiliation strings), treating each sample equally.
% These can be formally defined as follows:
% These metrics are straightforward for binary classification, with the formulas given as:

% However, these metrics require adjustments for multi-class classification to evaluate the performance of each class, as in the case of affiliation matching. 
In particular, the micro-averaged metrics are computed as follows:
{\footnotesize
\begin{align*}
\precision = \frac{\sum_{s} TP_s}{\sum_{s} (TP_s + FP_s)},\; 
\recall = \frac{\sum_{s} TP_s}{\sum_{s} (TP_s + FN_s)},\;
F_1 = \frac{2 \cdot \precision \cdot \recall}
{\precision + \recall}
\end{align*}
}

\noindent where $TP_s$ corresponds to the number of correct matches within the affiliation string $s$, $FP_s$ to the number of incorrect matches of $s$, and $FN_s$ to the matches of $s$ that were not found.
Finally, note that the larger the values of these measures, the better the effectiveness of the method under consideration.

For datasets where each affiliation string corresponds to a single organization identifier, and a prediction is always provided, we report accuracy, defined as:
{\footnotesize
\begin{align*}
% \text{Accuracy} = \frac{\text{TP}}{\text{TP + FP}}
\text{Accuracy} = \frac{Number\ of\ correct\ predictions}{Total\ number\ of\ predictions}
\end{align*}}

\subsubsection{Evaluation setting.}

We implemented \thiswork~in Python and released it as open-source software\footnote{\thiswork~GIT repository: \url{https://github.com/mkallipo/affiliation-matching}} under the Apache 2.0 license. 
The implementation uses the \texttt{unidecode}\footnote{\texttt{unidecode} library: \url{https://pypi.org/project/Unidecode/}} Python library for text normalization, and \texttt{scikit-learn}\footnote{\texttt{scikit-learn} library: \url{https://scikit-learn.org/}} for feature extraction and cosine similarity computation. 
\thiswork~includes three configurable parameters: 
$\texttt{sim\_o}$ and $\texttt{sim\_u}$, both ranging in the interval $(0, 1]$, and a $\texttt{window}$ parameter ranging from $1$ to $10$. 
An additional boolean flag, $\texttt{specific} \in \{\texttt{True}, \texttt{False}\}$, controls whether specific keywords like company names and acronyms are taken into account. For other methods, we used their publicly available GitHub implementations (see Section~\ref{sec:methods}).
Finally, experiments were conducted on an AWS M6g instance ($64$ CPU cores, $256$ GB RAM).

% $$Precision = \frac{C}{C + I},~~
% Recall = \frac{C}{C + U},$$

% $$F_{1} = 2 \cdot \frac{Precision \cdot Recall}{Precision + Recall }$$

% \noindent where $C$ denotes the number of correct matchings (true positives), $I$ the number of incorrect matchings (false positives), and $U$ the number of unmatched affiliations (false negatives).

% Notation:
% \begin{itemize}
% \item $C=$ Number of correct matchings (True positives)
% \item $I=$ Number of incorrect matchings (False positives) 
% \item $U=$ Number of unmatched affiliations (False Negatives) 
% \item $T=C+I+U$
% \item $\textbf{Return}=\frac{C+I}{T}$
% \item $\textbf{Precision}=\frac{C}{C+I}$
% \item $\textbf{Recall}=\frac{C}{C+U}$
% \item $\textbf{F}_1=2\cdot\frac{\mbox{Precision}\cdot \mbox{Recall}}{\mbox{Precision}+\mbox{Recall}}$

% \end{itemize}

\subsection{Configuring \thiswork}
\label{sec:eval-config}

In this section, we first present the configuration of our approach by tuning its parameters to maximize performance. Then, we vary the values of input parameters to further analyze their influence on the overall performance of \thiswork.

\subsubsection{Parameter tuning.}
\label{sec:params-tuning}

In this section, we evaluate the performance of \thiswork\ by optimizing its configuration parameters with respect to each of the evaluation metrics ($\precision$, $\recall$, and $\fone$) individually. To this end, we employ Bayesian optimization~\cite{bayesian} to efficiently explore the parameter space; we perform $1,000$  iterations per metric, using our algorithm as the objective function, set to maximize the corresponding measure in each case.

Specifically, the highest precision ($0.980$) was achieved with a $\texttt{window}$ size of $6$, $\texttt{sim\_o}$ set to $0.896$, $\texttt{sim\_u}$ to $0.898$, and $\texttt{specific}=\texttt{True}$. 
In contrast, the best recall ($0.917$) was obtained using a $\texttt{window}$ size of $3$, $\texttt{sim\_o} = 0.662$, $\texttt{sim\_u} = 0.313$, and $\texttt{specific}=\texttt{True}$. The optimal $\fone$ ($0.937$) was also achieved at $\texttt{window}$ size = 3, but with higher similarity thresholds ($\texttt{sim\_o} = 0.827$, $\texttt{sim\_u} = 0.426$) and again $\texttt{specific}=\texttt{True}$. 
It is evident that there is a trade-off between precision and recall depending on the similarity thresholds and the context-specific tuning of parameters.
% In the following section, we discuss each of these parameters affects the overall performance.

For clarity, we refer to the \thiswork\ variants that achieve the best precision, recall, and $\fone$ as $\thisp$, $\thisr$, and $\thisf$, respectively. These labels are used throughout the remainder of this paper to facilitate presentation and comparison of results.

\subsubsection{In-depth parameter analysis.}
\label{sec:in-depth-param-analysis}

\begin{figure}[h]
\centering
\begin{subfigure}[b]{0.328\textwidth}
\centering
\begin{tikzpicture}
\begin{axis}[
    xlabel={\texttt{window}},
    ylabel={Precision},
    label style={font=\scriptsize},
    xtick={1,...,10},
    xticklabel style={font=\tiny},
    ymin=0.97, ymax=0.985,
    ytick={0.97, 0.975, 0.98, 0.985}, % <- Specify your y-axis values here
    yticklabel style={font=\tiny, /pgf/number format/fixed, /pgf/number format/precision=3},
    width=4.4cm,
    height=4.4cm,
    grid=major,
]
\addplot[
    mark=*,
    mark size=2pt,
    color=black,
]
coordinates {
(1, 0.9754358161648178)
(2, 0.9760765550239234)
(3, 0.9790996784565916)
(4, 0.9790828640386162)
(5, 0.9798873692679002)
(6, 0.9798873692679002)
(7, 0.9798873692679002)
(8, 0.9798873692679002)
(9, 0.9798873692679002)
(10, 0.9798873692679002)
};
\end{axis}
\end{tikzpicture}
\caption{Precision}
\end{subfigure}%
\hfill
\begin{subfigure}[b]{0.328\textwidth}
\centering
\begin{tikzpicture}
\begin{axis}[
    xlabel={\texttt{window}},
    ylabel={Recall},
    label style={font=\scriptsize},
    xtick={1,...,10},
    xticklabel style={font=\tiny},
    ymin=0.899, ymax=0.92,
    ytick={0.895, 0.9, 0.905, 0.91, 0.915, 0.92}, % <- Specify your y-axis values here
    yticklabel style={font=\tiny,/pgf/number format/fixed, /pgf/number format/precision=3},
    width=4.4cm,
    height=4.4cm,
    grid=major,
]
\addplot[
    mark=*,
    mark size=2pt,
    color=black,
]
coordinates {
(1, 0.9002695417789758)
(2, 0.9117250673854448)
(3, 0.9171159029649596)
(4, 0.9164420485175202)
(5, 0.9164420485175202)
(6, 0.9164420485175202)
(7, 0.9164420485175202)
(8, 0.9164420485175202)
(9, 0.9164420485175202)
(10, 0.9164420485175202)
};
\end{axis}
\end{tikzpicture}
\caption{Recall}
\end{subfigure}%
\hfill
\begin{subfigure}[b]{0.328\textwidth}
\centering
\begin{tikzpicture}
\begin{axis}[
    xlabel={\texttt{window}},
    ylabel={$\fone$},
    label style={font=\scriptsize},
    xtick={1,...,10},
    xticklabel style={font=\tiny},
    ymin=0.92, ymax=0.94,
    yticklabel style={font=\tiny,/pgf/number format/fixed, /pgf/number format/precision=3},
    width=4.4cm,
    height=4.4cm,
    grid=major,
]
\addplot[
    mark=*,
    mark size=2pt,
    color=black,
]
coordinates {
(1, 0.9237199582027167)
(2, 0.9319916724496877)
(3, 0.9368932038834951)
(4, 0.9365244536940686)
(5, 0.9365244536940686)
(6, 0.9365244536940686)
(7, 0.9365244536940686)
(8, 0.9365244536940686)
(9, 0.9365244536940686)
(10, 0.9365244536940686)
};
\end{axis}
\end{tikzpicture}
\caption{$\fone$}
\end{subfigure}
\vspace{-0.2cm}
\caption{Performance metrics varying the $\texttt{window}$ parameter.}
\label{fig:window_size_subfigs}
\end{figure}
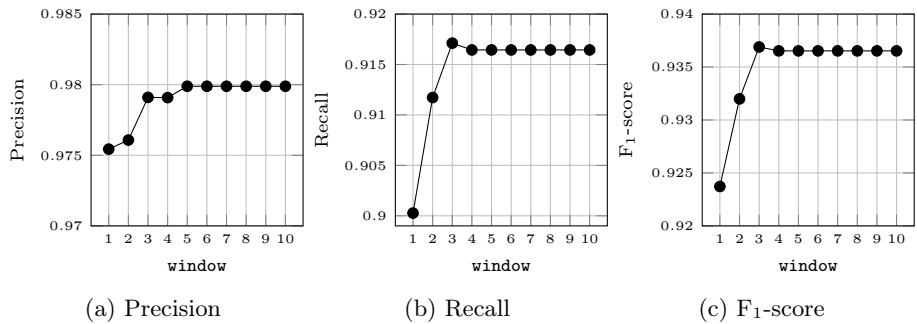

In this section, we discuss how each of the aforementioned parameters of \thiswork\ affects the performance.
Figure~\ref{fig:window_size_subfigs} illustrates the impact of varying the $\texttt{window}$ size (from $1$ to $10$) on precision, recall, and $\fone$ for our algorithm (see Section~\ref{sec:short}). 
The first plot shows that precision steadily improves when increasing $\texttt{window}$ size and reaches a plateau at a $\texttt{window}$ of $5$, maintaining the highest observed precision value of approximately $0.98$. 
Figure~\ref{fig:window_size_subfigs}b shows that recall peaks at $\texttt{window}$ $3$ (around $0.917$) and slightly declines beyond that, suggesting diminishing benefits for larger window sizes.
Figure~\ref{fig:window_size_subfigs}c reveals that $\fone$
also reaches its maximum at $\texttt{window}$ size $3$ ($0.937$), indicating the best balance between precision and recall occurs at this point.

\noindent

\vspace{-0.5cm}

\definecolor{purple}{HTML}{990099}  % Define the dark gray color

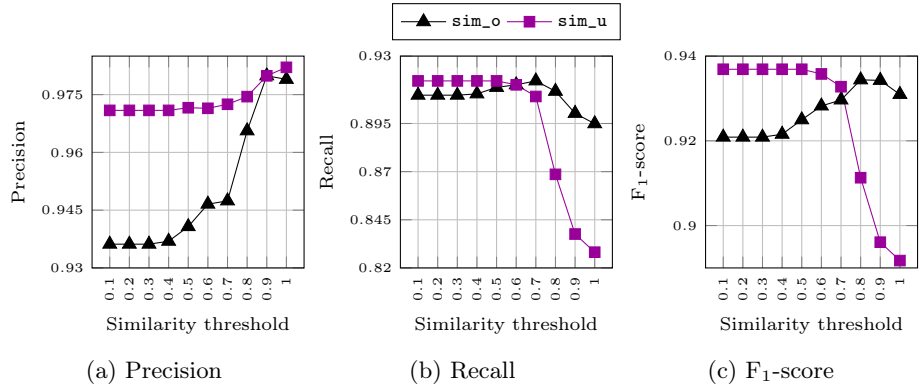
\begin{figure}[ht]
\centering

\begin{subfigure}[b]{0.328\textwidth}
\centering
\begin{tikzpicture}
\begin{axis}[
    xlabel={Similarity threshold},
    ylabel={Precision},
    label style={font=\scriptsize}, 
    xtick={0.1,0.2,...,1.0},
    xticklabel style={font=\tiny, rotate=90},
    ymin=0.93, ymax=0.985,
    ytick={0.93, 0.945,  0.96, 0.975},
    yticklabel style={font=\tiny, /pgf/number format/fixed, /pgf/number format/precision=3},
    width=4.4cm,
    height=4.4cm,
    grid=major,
    legend style={at={(0.5,-0.15)}, anchor=north, font=\scriptsize, legend columns=2} % Add common legend style here
]
\addplot[
    mark=triangle*, 
    mark size=3pt, 
    color=black
]
coordinates {
(0.1, 0.9361702127659575)
(0.2, 0.9361702127659575)
(0.3, 0.9361702127659575)
(0.4, 0.9369300911854104)
(0.5, 0.9407294832826748)
(0.6, 0.9465648854961832)
(0.7, 0.9474085365853658)
(0.8, 0.965598123534011)
(0.9, 0.9798873692679002)
(1.0, 0.9789644012944984)
};
\addplot[
    mark=square*, 
    mark size=2pt, 
    color=purple
]
coordinates {
(0.1, 0.9709090909090909)
(0.2, 0.9709090909090909)
(0.3, 0.9709090909090909)
(0.4, 0.9709090909090909)
(0.5, 0.9715950473415877)
(0.6, 0.9714494875549048)
(0.7, 0.9724907063197026)
(0.8, 0.974477958236659)
(0.9, 0.9798873692679002)
(1.0, 0.9820846905537459)
};
% Add the common legend entries here
% \legend{Method 1, Method 2}
\end{axis}
\end{tikzpicture}
\caption{Precision}
\end{subfigure}%
\hfill
\begin{subfigure}[b]{0.328\textwidth}
\centering
\begin{tikzpicture}
\begin{axis}[
    xlabel={Similarity threshold},
    ylabel={Recall},
    label style={font=\scriptsize}, 
    xtick={0.1,0.2,...,1.0},
    xticklabel style={font=\tiny, rotate=90},
    ymin=0.82, ymax=0.93,
    ytick={0.82, 0.845, 0.87, 0.895, 0.93},
    yticklabel style={font=\tiny, /pgf/number format/fixed, /pgf/number format/precision=3},
    width=4.4cm,
    height=4.4cm,
    grid=major,
    legend style={at={(0.5,1.25)}, anchor=north, font=\scriptsize, legend columns=2} 
]
\addplot[
    mark=triangle*, 
    mark size=3pt, 
    color=black
]
coordinates {
(0.1, 0.9097035040431267)
(0.2, 0.9097035040431267)
(0.3, 0.9097035040431267)
(0.4, 0.910377358490566)
(0.5, 0.9137466307277629)
(0.6, 0.9150943396226415)
(0.7, 0.9171159029649596)
(0.8, 0.9117250673854448)
(0.9, 0.9002695417789758)
(1.0, 0.894878706199461)
};
\addplot[
    mark=square*, 
    mark size=2pt, 
    color=purple
]
coordinates {
(0.1, 0.9171159029649596)
(0.2, 0.9171159029649596)
(0.3, 0.9171159029649596)
(0.4, 0.9171159029649596)
(0.5, 0.9171159029649596)
(0.6, 0.9150943396226415)
(0.7, 0.9090296495956873)
(0.8, 0.8685983827493261)
(0.9, 0.8376010781671159)
(1.0, 0.828167115902965)
};
\legend{$\texttt{sim\_o}$, $\texttt{sim\_u}$} % Add the common legend entries here
\end{axis}
\end{tikzpicture}
\caption{Recall}
\end{subfigure}%
\hfill
\begin{subfigure}[b]{0.328\textwidth}
\centering
\begin{tikzpicture}
\begin{axis}[
    xlabel={Similarity threshold},
    ylabel={$\fone$},
    label style={font=\scriptsize}, 
    xtick={0.1,0.2,...,1.0},
    xticklabel style={font=\tiny, rotate=90},
    ymin=0.89, ymax=0.94,
    yticklabel style={font=\tiny, /pgf/number format/fixed, /pgf/number format/precision=3},
    width=4.4cm,
    height=4.4cm,
    grid=major,
    legend style={at={(0.5,-0.15)}, anchor=north, font=\scriptsize, legend columns=2} 
]
\addplot[
    mark=triangle*, 
    mark size=3pt, 
    color=black
]
coordinates {
(0.1, 0.9208731241473398)
(0.2, 0.9208731241473398)
(0.3, 0.9208731241473398)
(0.4, 0.9215552523874488)
(0.5, 0.9249658935879945)
(0.6, 0.9282296650717704)
(0.7, 0.9296448087431693)
(0.8, 0.9343922651933702)
(0.9, 0.9342657342657343)
(1.0, 0.9309498773221172)
};
\addplot[
    mark=square*, 
    mark size=2pt, 
    color=purple
]
coordinates {
(0.1, 0.9368932038834951)
(0.2, 0.9368932038834951)
(0.3, 0.9368932038834951)
(0.4, 0.9368932038834951)
(0.5, 0.9368932038834951)
(0.6, 0.9357861853523082)
(0.7, 0.9327760362243122)
(0.8, 0.9112931955824723)
(0.9, 0.8960755813953489)
(1.0, 0.8917337234820776)
};
% \legend{Method 1, Method 2} % Add the common legend entries here
\end{axis}
\end{tikzpicture}
\caption{$\fone$}
\end{subfigure}

\vspace{-0.2cm}

\caption{Performance metrics varying similarity thresholds $\texttt{sim\_o}$ and $\texttt{sim\_u}$.}
\label{fig:sim-config}
\end{figure}

Figure~\ref{fig:sim-config} illustrates the effect of varying the similarity threshold parameters $\texttt{sim\_o}$ and $\texttt{sim\_u}$ on Precision, Recall, and $\fone$ (see Section~\ref{sec:cand}). 
Figure~\ref{fig:sim-config}a shows that precision increases consistently with higher values of $\texttt{sim\_o}$, starting at approximately $0.936$ for low values and peaking around $0.980$ at $\texttt{sim\_o} = 0.9$.
Similarly, higher values of $\texttt{sim\_u}$ achieve better precision, reaching a peak at $\texttt{sim\_u} = 1.0$. 
Figure~\ref{fig:sim-config}b shows that recall initially rises slightly and peaks at $\texttt{sim\_o} = 0.7$ (approximately $0.917$), after which it begins to decline.
In contrast, $\texttt{sim\_u}$ maintains a high (just below $0.92$) recall up to $\texttt{sim\_u} = 0.5$, but starts declining noticeably beyond that. 
Figure~\ref{fig:sim-config}c reflects the combined effect of these trends:
increasing $\texttt{sim\_o}$ improves $\fone$~steadily up to $\texttt{sim\_o} = 0.8$, reaching a maximum of around $0.934$, and then slightly decreases. This indicates that $\texttt{sim\_o}$ values around $0.7$ and $0.8$ offer the best overall performance.
On the other hand, $\texttt{sim\_u}$ follows a similar trend as in recall:  
it is stable and high at lower thresholds but steadily declines as $\texttt{sim\_u}$ increases (due to the respective drop in recall outweighing the gains in precision). 
This suggests that, while higher $\texttt{sim\_u}$ values improve precision, they do so at the cost of missing true positives, resulting in lower recall and ultimately a reduced $\fone$.

Finally, we evaluate the impact of the \texttt{specific} parameter (see Section~\ref{sec:filter}) on the best-performing variants of \thiswork\ with respect to precision, recall, and $\fone$ (denoted as $\thisp$, $\thisr$, and $\thisf$, respectively).
Table~\ref{tab:specific-results} shows that, across all configurations, the inclusion of the \texttt{specific} organizations yields consistent improvements in recall and $\fone$, while maintaining precision.
Specifically, $\thisp$~achieves the highest precision ($0.980$) both with and without \texttt{specific}, but its recall improves from $0.813$ to $0.821$ and F$_1$ from $0.889$ to $0.893$ when the \texttt{specific = True}. 
Similarly, $\thisr$ sees a recall increase from $0.908$ to $0.917$, that also results in an increase in $\fone$ from $0.925$ to $0.930$.
Lastly, $\thisf$, benefits from a recall gain ($0.903$ to $0.910$), resulting in an $\fone$ improvement from $0.933$ to $0.937$. 
These results showcase that \thiswork\ with the \texttt{specific} enabled is enhanced to capture relevant instances, especially in recall, without sacrificing precision.

\begin{table}[t]
    \centering
    \caption{Effect of the \texttt{specific} component of \thiswork.}
    \label{tab:specific-results}
    \begin{tabular}{ |l|ccc| }
    \hline
    \textbf{Approach} & \textbf{$\precision$} & \textbf{$\recall$}  & \textbf{$\fone$}\\
        \hline
            $\thisp$ w/o \texttt{specific}  &  $\bm{0.980}$  & 0.813 & 0.889  \\
            $\thisp$ & $\bm{0.980}$ & 0.821 & 0.893 \\
        \hline
            $\thisr$ w/o \texttt{specific}  &0.942  &0.908  &0.925  \\
            $\thisr$    &0.943 & $\bm{0.917}$ & 0.930 \\
        \hline
            $\thisf$ w/o \texttt{specific} &0.965 &0.903 & 0.933 \\
            $\thisf$  &0.965  &0.910  & $\bm{0.937}$\\
        \hline
    \end{tabular}
\end{table}

\subsection{Performance comparison with competitors}
\label{sec:eval-competitors}

This section evaluates the best-performing variants of \thiswork ($\thisp$, $\thisr$, and $\thisf$), against two competitor approaches: OpenAlex and S2AFF in two datasets (\db\ and Crossref).
% Additionally, for $S2AFF$, we consider two variants: 
% $S2AFF_{first}$ and $S2AFF_{all}$
% (see Section~\ref{sec:methods}).
The results are summarized in Table~\ref{tab:comparison-two-datasets}.

For \db, where each affiliation string is mapped to a list of organization identifiers, we report standard metrics (precision, recall, and $\fone$). 
Among all methods, $\thisp$ achieves the highest precision ($\precision = 0.980$), while $\thisr$ achieves the highest recall ($\recall = 0.917$). Importantly, $\thisf$ balances both metrics most effectively, reaching the best $\fone$ score of $0.937$, outperforming both S2AFF ($\fone = 0.901$) and OpenAlex ($\fone = 0.921$). 
% {\color{red}Notably, the $S2AFF_{all}$ variant emphasizes recall ($0.904$) but sacrifices precision substantially, resulting in the lowest $\fone$ among the methods compared.}

For Crossref, since only one ROR identifier is mapped to each affiliation string, 
we evaluate all algorithms over the 30,740 rows of the intersection of affiliation strings for which all methods returned a single result. 
To this end, we report the accuracy since it is a more suitable metric for this setup. 
Table~\ref{tab:comparison-two-datasets} shows that, $\thisf$ performs best, 
achieving the highest accuracy ($0.935$), 
slightly outperforming both OpenAlex ($0.920$) and S2AFF ($0.927$).
Notably, S2AFF performs comparatively better on this dataset than on \db, as it treats each affiliation string as representing a single affiliation, a strategy that better aligns with this dataset.
These results highlight that \thiswork~has superior precision and overall outperforms competition, with the $\thisf$ variant achieving the best $\fone$ across both datasets.

\begin{table}[t]
    \centering
    \caption{Evaluation against competitor approaches on \db\ and Crossref.}
    \label{tab:comparison-two-datasets}
    \begin{tabular}{|l|ccc|ccc|}
    \hline
    & \multicolumn{3}{c|}{\textbf{\db}} & \multicolumn{3}{c|}{\textbf{Crossref}} \\
    \cline{2-7}
    \textbf{Approach} & \textbf{$\precision$} & \textbf{$\recall$} & \textbf{$\fone$} & & \textbf{$\accuracy$} & \\
    \hline
    $\thisp$          & $\bm{0.980}$ & $0.821$ & $0.893$ &   & $-$ & \\
    $\thisr$          & $0.943$ & $0.917$ & $0.930$ &   & $-$ &  \\
    $\thisf$          & $0.965$ & $0.910$ & $\bm{0.937}$ &  & \bm{$0.935$} &  \\
    \hline
    $S2AFF$   & $0.964$ & $0.846$ & $0.901$ &  & $0.927$ & \\
    % $S2AFF_{all}$     & $0.615$ & $0.904$ & $0.732$ &  & $-$ & \\
    OpenAlex          & $0.914$ & $\bf{0.929}$ & $0.921$ &  & $0.920$ & \\
    \hline
    \end{tabular}
\end{table}

\section{Related work}
\label{sec:related}

Affiliation matching has become a prevalent task in metadata normalization, and several methods have been proposed to tackle it.
Among them, the OpenAlex institution parsing~\cite{openalex_institution_parsing,buttrick2023openalex,priem2022openalex} leverages an ensemble of two machine learning models trained on historical MAG data to classify and match institution strings. It further enhances performance by augmenting training data with synthetic affiliation strings.
S2AFF~\cite{kinney2023semantic} employs a three-stage pipeline: it first parses the input affiliation string into separate components (i.e., institution, country, address), 
it then retrieves the top candidate affiliations (using a retrieval index), and finally re-ranks them to produce the final result.
In addition, ROR~\cite{ror} added support to identify research organizations, by mapping free-text affiliation strings to ROR IDs, directly in their main API, using the most up-to-date version of the ROR registry.
While rather practical, it combines algorithmic logic with curated registry content, making it less transparent as a standalone algorithmic baseline.

Meanwhile, despite the importance of high-quality evaluation data, most existing datasets used for affiliation matching are only partially curated.
A recent dataset compiled by Crossref~\cite{tkaczyk2025crossref} includes a small subset of manually deposited organization identifiers (by Crossref members), 
but it primarily contains automatically identified mappings, 
which limits its reliability as a ground truth.
Additionally, it assigns a single organization identifier to each affiliation string, which is problematic in cases where the string refers to multiple organizations.
Furthermore, both the S2AFF Gold~\cite{s2aff,kinney2023semantic} and OpenAlex~\cite{openalex_test_dataset} datasets include some curated entries; for instance, OpenAlex incorporates entries curated by CWTS. 
However, in both cases, we observed inconsistencies and label noise, 
particularly in complex cases (see Section~\ref{sec:other-datasets} for details). 

\section{Conclusions}
\label{sec:conclusion}

In this work, we addressed the challenging problem of affiliation matching by introducing \thiswork, a novel approach for mapping raw affiliation strings to persistent organization identifiers. 
Our method leverages advanced parsing techniques to tackle the inherent complexity of affiliation strings, including cases where multiple organizations are mentioned. 
% {\color{blue}Since Affro does not require training, it eliminates the need for labeled data and retraining efforts. It is also easier to debug and refine by adjusting its parameters, allowing better control over false positives and false negatives. Finally, it remains stable over time, being less susceptible to data drift compared to machine learning models, though preprocessing rules may require occasional updates to handle new data patterns.}
We also compiled and made publicly available an expert-curated dataset designed to benchmark affiliation matching algorithms and provided a publicly accessible API to facilitate integration of our approach into third-party workflows. 
The evaluation of \thiswork~demonstrates its effectiveness in achieving higher performance than competitor approaches.
%In the future, we plan to exploit machine learning approaches to better identify the multiple organization names within affiliation strings, enhancing the accuracy of this aspect of our method.

% \begin{credits}
\section*{Acknowledgments} 
We summarize below the contributions of each author based on the CRediT taxonomy: 
\textbf{MK:} Conceptualization, Methodology, Software, Investigation, Data Curation, Writing - Original Draft, Visualization; 
\textbf{SC:} Conceptualization, Methodology, Investigation, Data Curation, Writing - Original Draft;
\textbf{MB:} Methodology, Investigation; 
\textbf{EA:} Data Curation; 
\textbf{PK:} Data Curation; 
\textbf{TV:} Conceptualization, Methodology, Investigation, Data Curation,  Writing - Original Draft, Supervision, Project administration, Funding acquisition

\vspace{1em}

\noindent This project has received funding from the European Union’s Horizon Europe framework programme under grant agreement No. 101058573. Views and opinions expressed are however those of the  author(s) only and do not necessarily reflect those of the European Union or the European Research Executive Agency. Neither the European Union nor the European Research Executive Agency can be held responsible for them.

\begin{figure}[!h]
     \centering
         \includegraphics[width=0.1\linewidth]{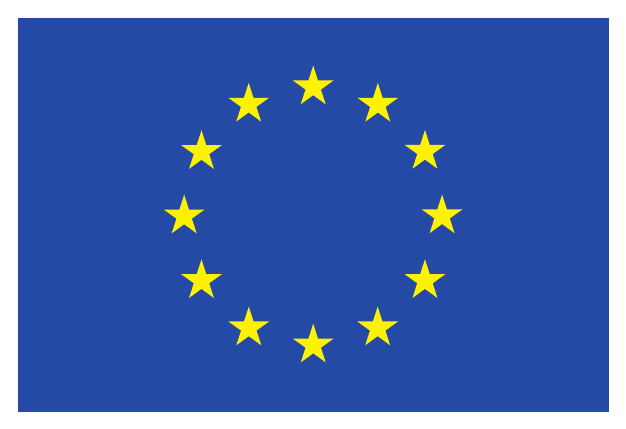}
\end{figure}
% \subsubsection*{\discintname}
% The authors declare that there are no competing interests.

% \end{credits}

\bibliographystyle{splncs04}
\bibliography{bibfile}
\end{document}